# The design study for a 500 MeV proton synchrotron with CSNS linac as an injector

Liang-Sheng Huang (黄良生)[1,2], Sheng Wang(王生)[1,2; 1)], Hong-Fei Ji (纪红飞)[1,2]

[1] Dongguan Campus, Institute of High Energy Physics (IHEP), Chinese Academy of Sciences (CAS)，Dongguan, 523803, China

[2] Dongguan Institute of Neutron Science (DINS), Dongguan, 523808, China

**ABSTRACT**

Using the China Spallation Neutron Source (CSNS) linac as the injector, a 500 MeV proton synchrotron is proposed for multidisciplinary application, such as biology, material and proton therapy. The synchrotron will deliver proton beam with energy from 80 MeV to 500 MeV. A compact lattice design was worked out, and all the important beam dynamics issues were investigated. The 80 MeV $H^-$ beam is stripped and injected into the synchrotron by using multi-turn injection. In order to continuously extraction the proton with small beam loss, the achromatic structure is proposed and slow extraction method with RF knock-out is adopted and optimized.

**Key words:** CSNS, proton synchrotron, multi-turn injection, slow extraction, RF knock-out

**PACS:** 41.85.ar, 29.27.ac.

## 1. Introduction

Molecular biology, single particle effect in electric components and cancer therapy by using proton beam has been highlighted worldwide for the past few years because of relative biology effectiveness and small beam diffraction. Many proton synchrotron complexes have being constructed in the world [1]-[5]. Using the China Spallation

[1)] wangs@ihep.ac.cn

Neutron Source (CSNS) [7] linac as the injector, a 500 MeV proton synchrotron is proposed for multidisciplinary application. It will deliver proton beam with energy from 80 MeV to 500 MeV. CSNS linac provide 80 MeV H$^-$ beam for injecting into a Rapid Cycling Synchrotron (RCS) with a repetition rate of 25 Hz and a duty factor of 1.25 %. The linac can provide an additional beam pulse between two injection beam pulses, so it can be an injector of another synchrotron.

A typical design of a proton synchrotron complex is to provide the proton beam with smooth spill, reliability and simplicity of operation. For low cost, the vertical betatron function should be small to decrease the aperture of the dipole magnet. To reduce the space charge effect and accumulate higher intensity beam, multi-turn injection is always used. The extraction is a key point. The quality of the extraction beam spill depends on the stable extraction from the synchrotron over the period of extraction. To share the linac beam with RCS, the transport line from CSNS linac to the 500 MeV synchrotron was designed. A compact 500 MeV synchrotron was worked out with optimized injection and extraction design, and the related beam dynamics studies were also done.

## 2. The design of transport line from linac to 500 MeV synchrotron

The CSNS LRBT (Linac to RCS Beam Transport) consists of a straight section and a bending section connected with a switch dipole. The H$^-$ beam is bended to the RCS or to beam dump controlled by the switch dipole as shown in Fig. 1 [8]. A port has been reserved for sharing linac beam in the beam dump line. To share the beam to the reserved port and keep the existing transport line unchanged, the H$^-$ beam is stripped to proton in the upstream of the switch dipole. As shown in Fig.1, a bump is generated by three pulse dipole magnets, and the short beam pulse for 500 MeV synchrotron is guided to the stripping foil and stripped. The proton beam will automatically guide to the reserved port while the dipole (LDSW) in beam dump line is switched off.

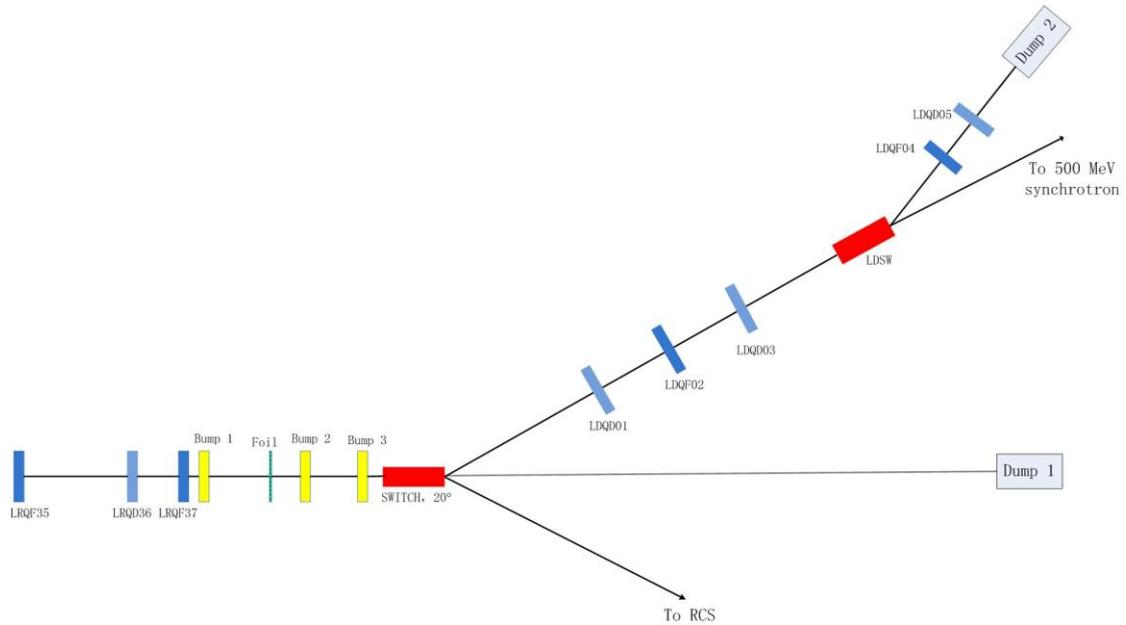

Fig. 1: The layout of transport line from linac to 500 MeV synchrotron

## 3. The lattice design of the synchrotron

A compact 500 MeV synchrotron is designed and its layout is shown in Fig. 2. The lattice of the synchrotron adopts the FODO cell based 2-fold structure, which consists of 8 dipoles and 12 quadrupoles powered by 4 families of power supply. The ring also makes up of two arc sections with reasonable dispersion function and two achromatic straight sections of 4 m respectively, which are reserved for the accommodation of the elements of injection, extraction, RF cavities and resonance sextupoles. The linac beam is injected into the ring by injection magnetic septum (IMS) and electrostatic infector (IEI). Two bump magnets are located in two sides of injection section with phase advance of 180 º for depressing closed orbit distortion (COD) in injection period. Accelerated beam is successively extracted by extraction electrostatic septum (EES) and magnetic septum (EMS). As a compact synchrotron, the regular dipole magnet with 1.8 m length and 1.6 T is used, and the good field region of the dipole is about 100 mm with gap of 60 mm. The aperture of the quadruple is generally 100 mm, but some of them in injection and extraction sections are bigger than 100 mm. Short drifts in the arc are reserved for chromatic sextupoles and RF kicker (RFKO).

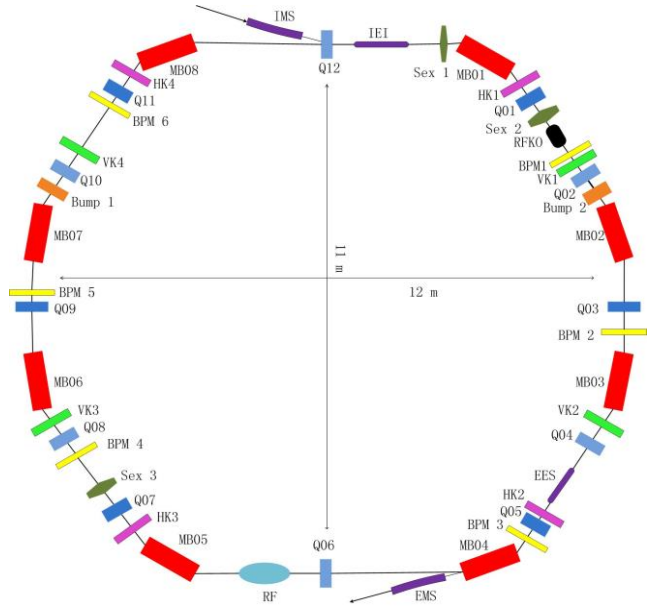

Fig. 2: The layout of 500 MeV synchrotron

The parameter of the synchrotron is listed in Table 1. The horizontal acceptance of the ring is 200 π.mm.mrad. Injection energy is designed to 80 MeV. Fig. 3 shows the twiss parameters of the synchrotron. The advantage of this design as follow: (1) it is a compact structure; (2) the horizontal and vertical betatron functions are limited 9 m; (3) the larger dispersion area along with smaller beta function; (4) there are two achromatic straight sections for robust injection and extraction. The COD is mainly from the field errors and misalignment of dipoles and quadrupoles. In order to correct COD, 6 horizontal BPMs, 6 vertical BPMs, 4 vertical steering magnets and 4 horizontal steering magnets are considered to meet the goal of 1 mm corrected COD.

Table 1: The main parameters of the synchrotron

| Parameters | Units | Values |
| --- | --- | --- |
| Circumference | m | 40 |
| Inj. Energy | MeV | 80 |
| Ext. Energy | MeV | 500 |
| Maximum $\beta_x/\beta_y$ (H/V) | m | 9.2/8.9 |
| Transition $\gamma$ | | 1.89 |
| Nominal Tunes (H/V) | | 1.71/1.24 |
| Natural Chromaticity (H/V) | | -0.47/-1.74 |
| Repetition Rate | Hz | 0.5 |
| Acceptance (H/V) | π.mm.mrad | 200/50 |
| Momentum Deviation | % | 0.3 |
| Accumulated Proton Number | | $5*10^{11}$ |
| Harmonic Number | | 1 |
| RF frequency | MHz | 2.8~5.8 |

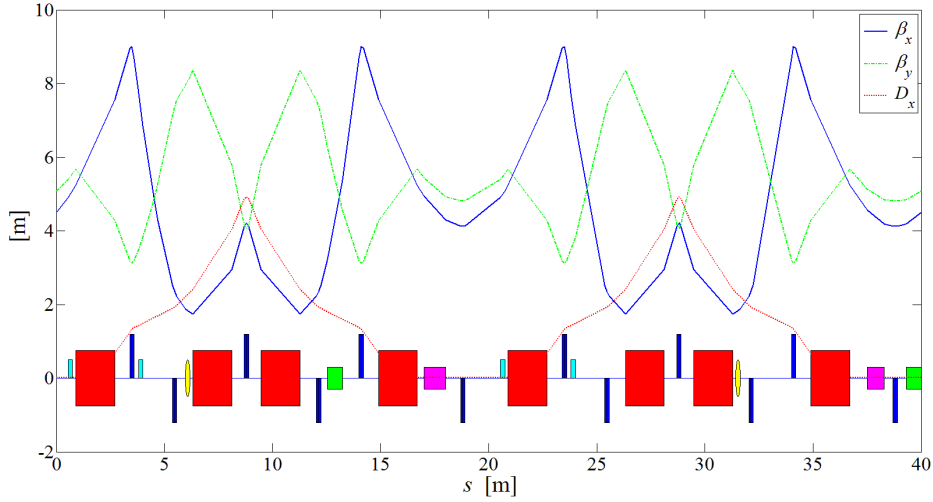

Fig. 3: The twiss parameters of the synchrotron

## 4. Injection design and simulation

80 MeV proton beam from CSNS linac is injected into the synchrotron by using multi-turn injection in horizontal plane. The injection system consists of IMS, IEI and two bump magnets (bump 1, bump 2) with phase advance of 180 º. A turn number for each injection period is about 10, and the number of injected particle per turn is more than $8 \times 10^{10}$ for the typical injector beam. For proton beam with energy of 80 MeV, when the space charge limited intensity is limit to $5 \times 10^{11}$ ppp, the space charge tune shift is much smaller than 0.1. Injection beam is painted by sweeping the bump height.

A code for injection tracking of the synchrotron was developed base on AT [9][10] and the beam injection was studied by the code. The linac beam of 1000 macro-particles per turn is injected in the tracking, a constant ramping rate of the bump field is assumed, and the thickness of IEI is set to 0.2 mm. The beam distribution in horizontal phase space after painting is shown in the left of Fig. 4. The green line is the location of IEI, the red dot denotes the injection beam and blue dot is accumulated cycling beam. Injection beam is painted around about 60 π.mm.mrad in horizontal plane. The survival rate of injection beam after 40 turns is displayed in right plot of Fig. 4. The efficiency of injection without space charge effect is about 60 % and the bean loss at IEI is about 20 %.

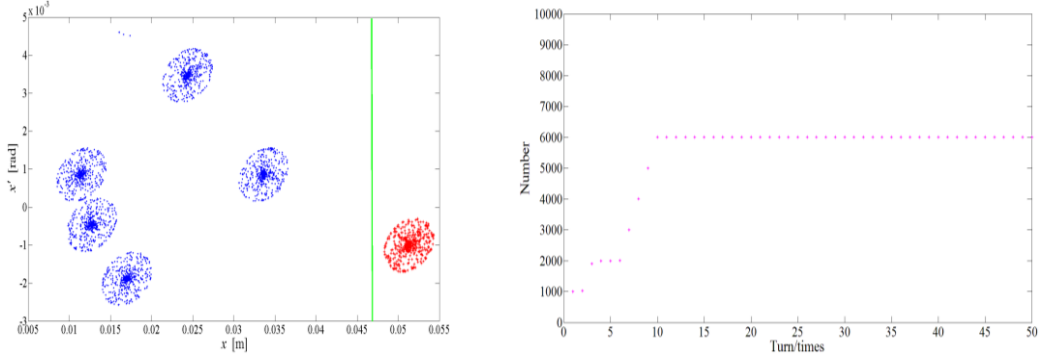

Fig. 4: The beam distribution at injection point (left) and accumulate rate (right)

## 5. Extraction design and simulation

The 3$^{st}$ resonance slow extraction with RF knock-out scheme is used to obtain continuous and smooth spill. The layout is shown in Fig. 2. Two resonance sextupoles (Sex 1 and Sex 3), RFKO, EES and EMS were used. Resonance sextupoles is placed in two dispersion-free regions. The phase advance between the main resonance sextupole (Sex 1) and EES is well above the permissible minimum of 217 ° and close to the ideal value of 225 °. The phase advance from EES to EMS is 39 °, which can provides 62 % kick of that of 90 ° phase advance. The RFKO is located in arc. The dispersion function is correctly configured with $D_x > 0$ and $D_x' < 0$ at EES for meeting the Hardt condition [11]. At the resonance sextupole, $\beta_x \approx \beta_y$ induce the large coupling, but this is not a problem for beam stable in vertical plane. $\beta_x$ is large to enhance the kick at EES and $\beta_y$ is small to reduce the aperture requirement of EMS.

The accelerated beam is extracted by slow extraction scheme using a third order resonance $\nu_x = 5/3$ with transverse RFKO. The tune in horizontal plane is adjusted from 1.71 to 1.67 before the beam extraction. The phase space is shaped into a triangle when resonance sextupoles operate. Particles inside the triangle are stable while particles out of the triangle are unstable and it will be extracted quickly. Stabilities of extracted beam parameters, for example beam position, beam size and momentum deviation, are very important issues for special experiment, such as spot scanning irradiation and single particle effect study. In the RFKO method, the separatrix is kept constant during the whole extraction period, so it is easy to control the extraction by using RFKO.

The simulation code based on AT [10] has also been developed in order to analyze a mechanism of realistic RFKO and to realize more precise control. The maximum spiral step is bigger than 4.5 mm. A few percent of beam loses at EES and the extraction efficiency is over 95 % in the tracking. The two extraction schemes, the scheme with sextupoles only and the scheme with sextupoles and RFKO, are compared. Fig. 5 shows the extraction beams versus extraction turn, the left plot is the extraction with only resonance sextupoles, the particle is extracted about some hundred turns, and a lot of particles inside of stable triangle are difficult to be extracted, while the particles can be continuously extracted by combing the sextupole and RFKO, as shown in the right plot, and the spill is more smooth.

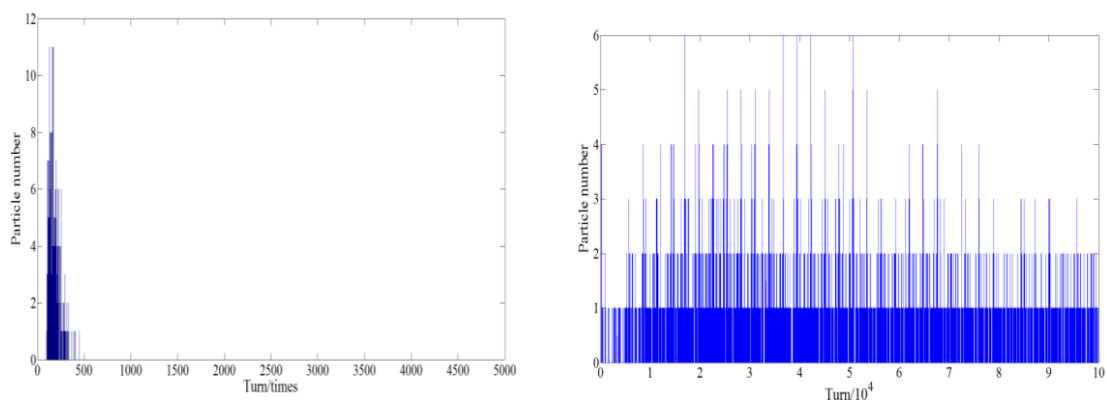

Fig. 5: Extraction particle number per turn by using only the resonance sextupole (left) and the sextupole with RFKO (right)

The ripple of the power supply of dipoles and quadrupoles induces fluctuation of tune, which changes the stable triangle, thus the number of extraction particles waves with the ripple of the power supply, which is awful for spot scanning, so the ripple should be well controlled. The ripples effects were studied by simulation with different ripple level. For a few percent of extraction variation, the ripples of dipoles and quadrupoles are 5 ppm and 20 ppm, respectively.

The orbit of extraction beam passes through EES and EMS is shown in Fig. 6. To reduce the beam loss in extraction, the apertures of the dipole and two quadrupoles are bigger than other general magnets in the ring. The electric intensity of EES and magnetic field of EMS are 6.1 kV/mm and 0.98 T.

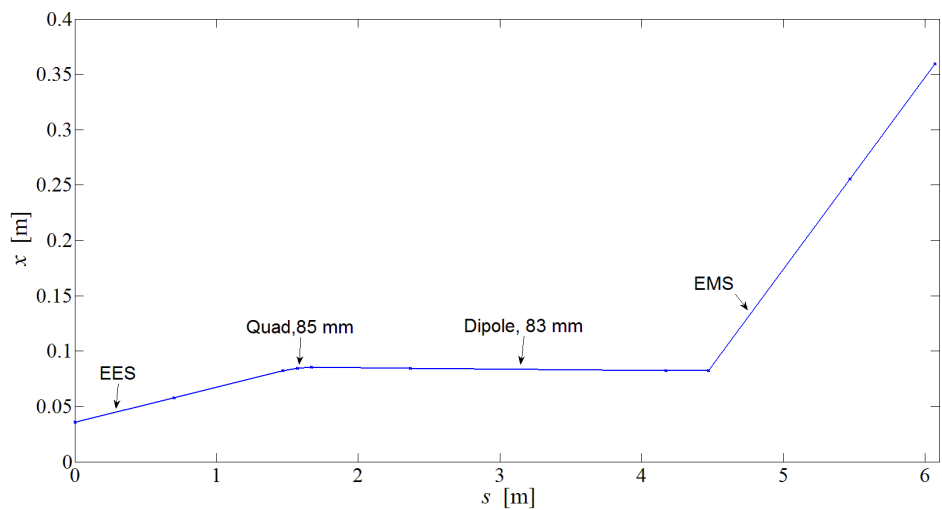

Fig. 6: The orbit of extraction beam. The value of magnets is the half aperture

## 6. Summary

Using the China Spallation Neutron Source (CSNS) linac as the injector, a 500 MeV proton synchrotron is proposed for multidisciplinary application. The compact synchrotron with two achromatic straight sections is useful to robust injection and extraction. The acceleration-driven resonant extraction scheme using RFKO is chosen to offer the smoothest spill of the particles to particular experiment. Based on AT, the simulation code of injection and extraction is developed. The particle tracking simulation for injection confirms the required conditions of bump magnets, IMS and IEI. The particle tracking simulation for extraction shows that the beam can be extracted with high efficiency. The influence of ripple of magnets and the orbit of extracted beam are optimized for the smooth spill.

## Acknowledgements

We would like to acknowledge the support of many colleagues, especially Prof. N Huang for many discussions and comments in the study.